\documentclass[aps,prb,superscriptaddress,showpacs,twocolumn]{revtex4}
\usepackage{graphicx,amsmath, amssymb, pifont}
\usepackage[usenames]{color}

\begin{document}

\title{
Doppler Shift in Andreev Reflection from a Moving Superconducting Condensate in Nb/InAs Josephson
Junctions}

\author{F.~Rohlfing}
\affiliation{Institute for Exp.~and Applied Physics, Universit\"{a}t
Regensburg, D-93040 Regensburg, Germany}
\author{G.~Tkachov}\altaffiliation[Present address: ]{Institute for Theoretical Physics and Astrophysics, University of W\"urzburg, Germany}
\affiliation{Institute for Theoretical Physics,
Universit\"{a}t Regensburg, D-93040 Regensburg, Germany}
\author{F.~Otto}\affiliation{Institute for Exp.~and Applied Physics, Universit\"{a}t
Regensburg, D-93040 Regensburg, Germany}
\author{K.~Richter}\affiliation{Institute for Theoretical Physics,
Universit\"{a}t Regensburg, D-93040 Regensburg, Germany}
\author{D.~Weiss}\affiliation{Institute for Exp.~and Applied Physics, Universit\"{a}t
Regensburg, D-93040 Regensburg, Germany}
\author{G.~Borghs}\affiliation{Interuniversity Micro-Electronics Center (IMEC),
Kapeldreef 75a, B-3001 Leuven, Belgium}

\author{C.~Strunk}\affiliation{Institute for Exp.~and Applied Physics, Universit\"{a}t
Regensburg, D-93040 Regensburg, Germany}

\begin{abstract}
We study narrow ballistic Josephson weak links in a InAs quantum
wells contacted by Nb electrodes and find a dramatic magnetic-field
suppression of the Andreev reflection amplitude, which occurs even
for in-plane field orientation with essentially no magnetic flux
through the junction. Our observations demonstrate the presence of a
Doppler shift in the energy of the Andreev levels, which results
from diamagnetic screening currents in the hybrid Nb/InAs-banks. The
data for conductance, excess and critical currents can be
consistently explained in terms of the sample geometry and the
McMillan energy, characterizing the transparency of the
Nb/InAs-interface.
\end{abstract}


\maketitle In recent years a detailed microscopic understanding of
the proximity effect has emerged. There is now agreement that in
highly transparent Josephson junctions formed by a metallic weak
link in good contact with two superconducting  (SC) banks the
supercurrent is carried by Andreev bound states (ABS)~\cite{kulik}.
These come in pairs corresponding to the opposite directions of
Cooper-pair transfer mediated by multiple Andreev reflection (MAR)
at the SC/metal interfaces provided that the acquired
quasi-particle phase is a multiple of $2\pi$.\cite{schapers} 
At currents exceeding the critical current $I_C(T,B)$ MAR between the
SC banks manifests itself in the current-voltage
characteristics as subharmonic gap structures at voltages
$eV_n=2\Delta/n$, where $\Delta$ is the SC energy gap and
$n=1,2,...$. At higher voltage $eV\gg 2\Delta$, $I(V)$ becomes
linear with an excess current $I_\textrm{exc}=I(V)-G_NV$ determined
by a single Andreev reflection (AR) probability $|a(\varepsilon)|^2$
($G_N$ is the normal state conductance).~\cite{SGS,BTK}

Weak links formed by a two-dimensional electron gas (2DEG) in 
semiconductor quantum wells~\cite{vWees} are of particular interest
because here the ballistic transport can be studied. In very
high magnetic field perpendicular to the 2DEG, theory~\cite{hoppe}
and experiments~\cite{eroms,batov} have demonstrated Andreev
transport via edge states. Indirect evidence for a strong
magnetic-field effect on AR was experimentally found in antidot
billiards.\cite{eroms2} The case of parallel field, with respect to
the 2DEG, is equally intriguing: as ideally no magnetic flux threads
the 2DEG, one may naively expect the Josephson current to survive up
to the critical fields of the SC leads. This is not the case, but
the underlying mechanism of the supercurrent suppression is still
unclear. This question, also relevant for other 2D hybrid
systems,~\cite{heersche} is among the issues this study focuses on.

In this Communication, Nb/InAs Josephson junctions of different
width are studied in a four-terminal lead configuration within the
2DEG. This allows us to separately determine the transparencies of
the InAs weak link and Nb/InAs interfaces, and identify an
additional energy scale in the electronic spectrum of the
hybrid SC terminals. We observe a very strong suppression of both
the AR probability and supercurrent in weak magnetic fields of 4 and
100~mT for perpendicular and parallel orientations, respectively.
This unexpectedly rapid decay can be traced back to diamagnetic
supercurrents in Nb, which break the time-reversal symmetry of Andreev reflection.

\begin{figure}[t]
\includegraphics[width=75mm]{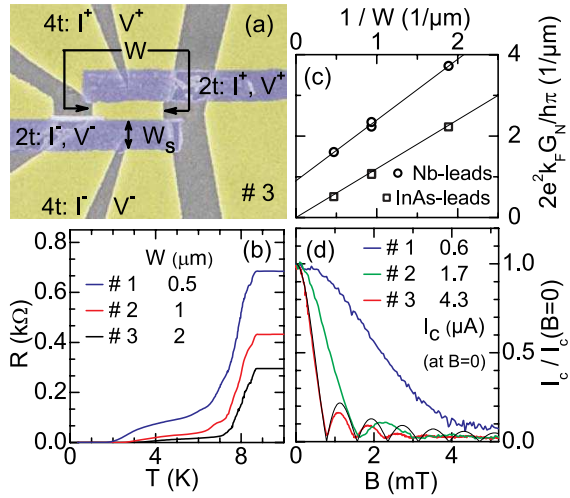}
\caption{(color online) (a) Scanning electron micrograph of sample \#3 with width
$W=2\,\mu$m and length $L=0.6\,\mu$m. Regions with InAs
two-dimensional electron gas are shown in yellow. Etched regions are
light grey. The horizontal dark grey stripes are made from a 35~nm
thick Nb.  (b) $R(T)$ for three samples with width
$W=0.5,\,1,\,2\,\mu$m. (c) Resistance vs.~$1/W$ at 10~K for the
4t-configuration with InAs leads ($\textsf{O}$) and the
2t-configuration with Nb leads ($\square$). (d) 
Normalized supercurrent $I_C$ vs. perpendicular magnetic field
$B_\perp$ for different widths $W$. The black curve is the
standard Fraunhofer pattern fitting the data for sample \#3
($W=2\,\mu$m).}\label{sample}
\end{figure}

The samples were prepared from a heterostructure containing an
InAs-quantum well with thickness $d_N=15\,$nm and a mean free path
around 3.7~$\mu$m, confined between two AlGaSb layers. The electron
density was $7.8\cdot10^{15}\,$m$^{-2}$, resulting in a Fermi
wavelength $\lambda_F=2\pi/k_F=28\,$nm.~\cite{eroms} First, a
constriction of width $W$ with a four-terminal (4t) lead structure
(shown in yellow in Fig.~\ref{sample}a) was patterned using electron
beam lithograpy (EBL) and etched by reactive ion etching in a
SiCl$_4$-plasma. In a second EBL step two Nb stripes were deposited
onto the InAs (horizontal grey bars in Fig.~\ref{sample}a) after
removal of the top AlGaSb layer and in-situ Ar-ion cleaning of the
InAs surface. The Nb-stripes have a width $W_S=1\,\mu$m and a
thickness $d_S=35\,$nm; their distance $L=600\,$nm defines the
length $L$ of the junction. This results in a Thouless energy of
$\varepsilon_{Th}=\hbar v_F/L=0.8\,$meV $\lesssim \Delta$ for all
samples, $v_F$ being the Fermi velocity in the normal metal. We have
prepared four samples on two chips in the same batch. The chips
contained junctions with $W=0.5$ and $1\,\mu$m (denoted as \#1 and
\#2a), and $W=1$ and $2\,\mu$m (denoted \#2b and \#3), respectively.

Figure~\ref{sample}b shows the two-terminal (2t) resistance of
samples \#1, \#2a, and \#3 measured through the Nb stripes
vs.~temperature $T$. The resistance drops at the transition
temperature of the Nb stripes around $8.3\,$K and again between 2
and $3\,$K, where the constriction becomes superconducting. In
Fig.~\ref{sample}c we plot the 2t-resistance measured across the Nb
stripes and the 4t-resistance measured within the InAs layer at
$T=10\,$K as a function of $W$. The 4t-resistance scales as $1/W$,
obeying the ballistic Landauer-B\"{u}ttiker formula:
\begin{equation}
\label{Eq_sharvin}G_N=\frac{2e^2}{h}\,\sum_n\,{\cal
T}_n=\frac{2e^2}{h}\frac{k_FW}{\pi}\cdot \langle{\cal T}\rangle\;,
\end{equation}
where $\langle{\cal T}\rangle$ is the average channel transparency.
From the slope of $G_N(W)$ we extract  $\langle{\cal T}\rangle=0.8$.
The 2t-resistance at $10\,$K is also proportional to $1/W$, however,
with an offset caused by the normal state resistance of the
Nb-stripes and the Nb/InAs-interface resistance. Finally,
we show in Fig.~\ref{sample}d the interference pattern of the
junctions in a transversal magnetic field.
For the widest junction the critical current $I_C(B)$  exhibits the expected Fraunhofer
pattern.\cite{tinkham} For the narrower ones the higher order
maxima are suppressed, and in the $0.5\;\mu$m-wide device $I_C(B)$
resembles more a Gaussian rather than a Fraunhofer pattern.
This will be addressed below.

\begin{figure}[t]
\includegraphics[width=75mm]{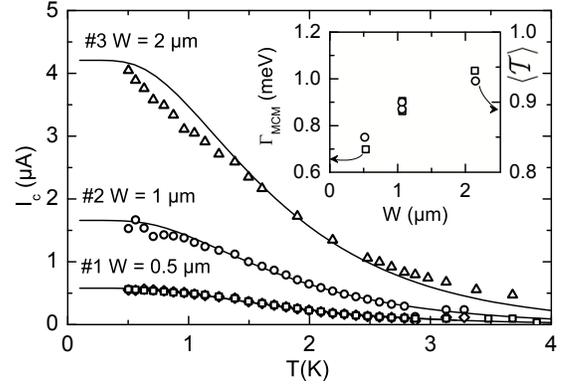}
\caption{Critical current vs.~temperature $T$ for different widths
$W$ together with best fits according to our model. Inset: Values of
$\langle{\cal T}\rangle\;(\textsf{O})$ and $\Gamma_{McM}\;(\square)$
resulting from the fits.}\label{IC_T}
\end{figure}

Information about the spectral distribution of ABS is contained in
the $T$-dependence of the supercurrent. In Fig.~\ref{IC_T} we
present the critical current $I_C$ at $B=0$ as a function of
temperature for different widths $W=0.5,\,1$ and $2\,\mu$m of the
InAs constriction. With decreasing temperature $I_C(T)$ initially
increases and then saturates at $T\lesssim1\,$K. The solid lines are
fits based on the approach of Grajcar et al.~\cite{grajcar} who
adapted the scattering theory for the Josephson
current~\cite{brouwbeen} to 2D ballistic proximity structures. Two
parameters were used in these fits: the average normal-state
transmission probability $\langle{\cal T}\rangle$ and the McMillan
energy $\Gamma_{McM}=\tau_{SN}\cdot\hbar v_F/2d_N,$~\cite{mcmillan}
which is the Thouless energy of the InAs quantum well multiplied by
the Nb/InAs interface transparency $\tau_{SN}$. $\Gamma_{McM}$
determines the strength of the proximity-induced superconducting
correlations in the InAs regions underneath the Nb film, which act
as effective superconducting terminals described by the Green's
functions of the McMillan model.\cite{grajcar,golubov} Clearly, the
smaller $\tau_{SN}$, the weaker the proximity effect. Because
$\langle{\cal T}\rangle$ influences mainly the saturation value of
$I_C(T\rightarrow0)$, while $\Gamma_{McM}$ determines the decay of
$I_C$ with $T$, these two parameters can be extracted independently
from the measured $I_C(T)$.

\begin{figure}[t]
\includegraphics[width=75mm]{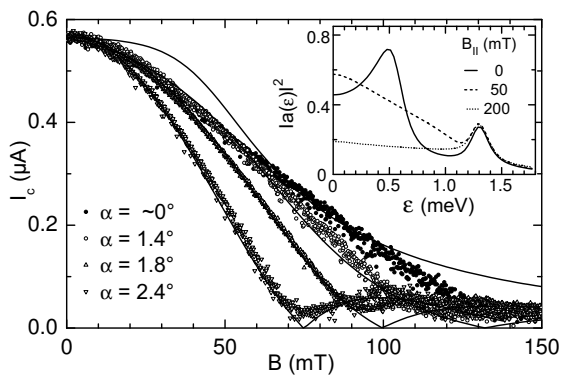}
\caption{Critical current vs.~$B$ for sample \#1 
at small angles $\alpha$ between ${\bf B}$ and 2DEG. 
The lines are theoretical fits (see text). The inset shows the double
peak structure of $|a(\varepsilon)|^2$ characteristic of hybrid SC
and its suppression by $B_{||}$.}\label{IC_B}
\end{figure}

The fit parameters are given in the inset of Fig.~\ref{IC_T}. The values of
$\langle{\cal T}\rangle$ scatter by $\pm5\,\%$ around 0.9, which is slightly higher, but still
close to $\langle{\cal T}\rangle=0.8$ estimated independently from 4t-resistance
(\ref{Eq_sharvin}). The values of $\Gamma_{McM}$ scatter by $\pm 15\,\%$ around $0.9\,$meV.
From this, we find $\tau_{NS}\approx 0.06$ which is much smaller than
$\langle{\cal T}\rangle$. The high mobility of the InAs-quantum well provides the high transparency
of the constriction, while the comparatively low Nb/InAs interface transparency is sufficient to
transform the Nb/InAs-bilayer into an effective superconductor with the additional energy scale
$\Gamma_{McM}$ (see also inset in Fig.~\ref{IC_B}).

\begin{figure}[t]
\includegraphics[width=73mm]{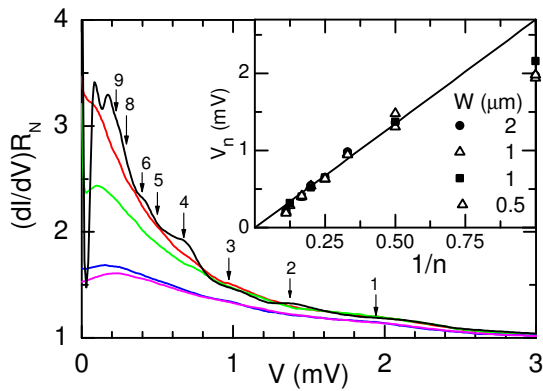}
\caption{(color online) Measured differential conductance $dI/dV$ of
sample \#2a ($W=1\mu$m) for different values of a perpendicular
magnetic field $B_\perp=0,4,10,40,$ and 100~mT, from top to bottom.
Arrows indicate the subharmonic gap structure induced by multiple
Andreev reflections. Inset: Voltages of the subharmonic gap
structures for all samples. The solid line indicates the expected
slope for $\Delta_{Nb}=1.35\;$meV. }\label{MAR}
\end{figure}

Having all sample parameters fixed, we now turn to the main topic of
our study, i.e. the magnetic field dependence of Andreev reflection.
First, we examine $IV$- and $dI/dV$-characteristics.
Figure~\ref{MAR} shows the differential conductance $dI/dV$ of
sample \#2a ($W=1\,\mu$m) for different values of perpendicular field 
$B_\perp$. At $B_\perp=0$ subharmonic gap structures (see arrows) appear 
at integer fractions $V_n=2\Delta_{Nb}/ne$ of $2\Delta_{Nb}$. 
The location of the steps vs. $1/n$, plotted in the inset of Fig.~\ref{MAR},  
agrees very well with the value of $\Delta_{Nb}=1.35\;$meV. 
The $n=1$ step appears at lower voltage as predicted in
Ref.~\onlinecite{samuelson}. A tiny perpendicular field of
$4\,$mT washes out these features, vanishing completely at $10\,$mT.
Also, the enhancement of $dI/dV$ at low bias is suppressed. 
It is known~\cite{merkt,schapers} that the subharmonic
gap structure and low bias enhancement of $dI/dV$ originate from
multiple AR at finite $V$. The large number of steps
independently confirms the high AR probability $|a|^2$ expected from
the high values of $\langle{\cal T}\rangle$. The rapid smearing of
these steps thus indicates a surprisingly strong magnetic field
dependence of $|a|^2$.

\begin{figure}[t]
\includegraphics[width=80mm]{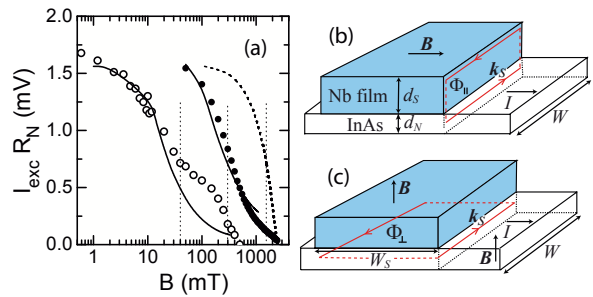}
\caption{(color online) (a) Measured excess current of sample \#2a
vs. perpendicular ($\circ$) and parallel ($\bullet$) magnetic field.
The solid lines result from our model including the Doppler shift
$\varepsilon_D(k_S)$ (see text) and the dashed line shows
$I_\textrm{exc}$ due to the suppression of $\Delta_{Nb}(B_\|)$
only.\cite{skalski} (b) and (c) One of the Nb/InAs terminals in
parallel and perpendicular fields. Red arrows indicate the SC
order-parameter phase gradient, ${\bf k}_S=\nabla\gamma$, with
$k_S=-\pi/W\times\Phi/\Phi_0 $ determined by the magnetic flux
$\Phi$ through the bilayer for given field orientation.
}\label{Iexc}
\end{figure}

This observation is further substantiated by the behavior of the
excess current $I_\textrm{exc}$,~\cite{BTK} obtained by
integrating $dI/dV-G_N$ (see Fig.~\ref{MAR}) over voltage. 
Since $I_\textrm{exc}$ results from a single AR at $V>2\Delta_{Nb}$, 
it is more robust than MAR whose amplitude is $\propto |a|^{2n}$. 
Figure~\ref{Iexc} shows $I_\textrm{exc}$ on a
logarithmic scale for both field orientations. It is
suppressed to $\sim 50$~\% at $B_\perp=30\;$mT and
$B_\parallel=300\;$mT. The field $B_\parallel$ exceeds $B_\perp$ by
an order of magnitude, but is still a factor of 5 smaller than
needed for an appreciable reduction of the gap $\Delta_{Nb}(B)$ as
follows from the pair-breaking theory~\cite{skalski} (dashed line in
Fig.~\ref{Iexc}).

We attribute the behavior of $I_\textrm{exc}(B)$ to the magnetic
field suppression of the Andreev reflection in the InAs proximity
regions underneath the Nb films. In the presence of the external
field ${\bf B}=\nabla\times{\bf A}({\bf r})$, the SC order parameter
acquires an inhomogeneous phase, $\gamma({\bf r})=-
2\pi\Phi^{-1}_0\int^{\bf r}_{{\bf r}_0} {\bf A}({\bf r}^\prime)d{\bf
r}^\prime$, whose gradient ${\bf k}_S=\nabla\gamma$ induces
diamagnetic currents in the entire Nb/InAs structure, including the
InAs proximity layer (see, Fig.~\ref{Iexc}b and c). If ${\bf k}_i$
and ${\bf k}_f$ denote the wave vectors of initial and final
quasi-particle states in the AR-process, momentum conservation
requires ${\bf k}_i+{\bf k}_f={\bf k}_S$, since the Cooper-pairs
absorbed by the moving SC condensate have momentum $\hbar{\bf k}_S$.
This leads to a {\it Doppler shift} $\varepsilon_D=-\hbar^2k_\perp
k_S/2m^*$ of the energy of the AR quasi-particle ($k_\perp$ is the
transverse wave number).~\cite{gt} The inset in Fig.~\ref{IC_B}
illustrates the energy dependence of $|a(\varepsilon)|^2$. The fact
that each of the  Nb/InAs-banks constitutes {\it a single} hybrid
superconducting terminal is illustrated the two BTK-like peaks in
the spectum, one at $\Delta_{Nb}$ and another at a smaller energy
determined by $\Gamma_\text{MCM}$. Similar hybrid excitation spectra
have recently been observed by scanning tunneling spectroscopy in
diffusive Al/Cu-structures \cite{saclay}. The suppression of
$|a(\varepsilon)|^2$ occurs in the range $|\varepsilon|\lesssim
\Gamma_{McM}$, when $\varepsilon_D$ becomes comparable to the
McMillan energy $\Gamma_{McM}$.\cite{note2}

The solid lines in Fig.~\ref{Iexc}a show $I_\textrm{exc}(B)$
obtained from the BTK-model,\cite{BTK} in which we express the AR
amplitude $a(\varepsilon)$ in terms of the McMillan's Green's
functions~\cite{grajcar,golubov} accounting for the Doppler shift
$\varepsilon_D$ in the InAs proximity layer. Using the parameters
{\it extracted from Fig}.~\ref{IC_T}, the model reproduces the
measured curves quite well (except for the shoulder in
$I_\textrm{exc}(B_\perp)$, which at present is not understood). The
model accounts for the ratio of the $B_\perp$  and $B_\|$ scales.
This has a simple interpretation if we notice that the relevant
value of $k_S$ is related, via the circulation theorem, to the
magnetic flux $\Phi$ threading the Nb/InAs bilayer for given field
orientation (Fig.~\ref{Iexc}b and c): $k_S=-\pi/W\times
\Phi/\Phi_0$. Neglecting the field inhomogeneity in
Nb,\cite{conditions} we find $\Phi=B_\| W(d_S+d_N)$ and
$\Phi=B_\perp WW_S$ for the parallel and perpendicular cases,
respectively. Consequently, the expected field ratio is
$B_\|/B_\perp =W_S/(d_S+d_N)=20$, which is within the margins of the
experimental uncertainty.

The dramatic reduction of the AR amplitude $a(B)$, inferred from the
behavior of $I_\textrm{exc}$, can also explain the suppression of
the higher lobes of the critical current in the Fraunhofer pattern
in weak perpendicular magnetic field (see Fig.~\ref{sample}d), since
the same $a(B)$ determines the energies of the ABS and hence
$I_C(B)$.\cite{brouwbeen,grajcar} We also note that the smaller the
channel width $W$ is, the higher are the values of $B_\perp$
required to reach $\Phi=\Phi_0$ in the junction and the stronger is
the suppression of $I_C(B)$ at this point.

Finally, we examined the critical current $I_C$ in
a parallel magnetic field for samples \#1 and \#2a. Again,
even for the perfectly parallel field there is the phase gradient
due to the vector potential (Fig.~\ref{Iexc}b) that
induces a Doppler energy shift of the ABS. When it becomes
comparable to $\Gamma_{McM}$, counter-propagating ABS overlap in
energy, and their contributions to $I_C$ gradually cancel. 
Similar to $I_\textrm{exc}$, the suppression of $I_C$
occurs at $\sim 100$~mT. In Fig.~\ref{IC_B} we plot $I_C(B)$ for sample \#1 
for several almost parallel field orientations. The small perpendicular
component of ${\bf B}$ still leads to a Fraunhofer-like minimum in
$I_C$, which allows a precise determination of the angle $\alpha$
between ${\bf B}$ and the junction plane. This can be described by
$I_c(B)=I_c(B_\|)\left| \sin(\pi A B_\perp/\Phi_0
)/(\pi A B_\perp/\Phi_0) \right|$, where $B_\perp =B\sin\alpha$,
$B_\| =B\cos\alpha$ and $A$ is the junction area. Unlike the
standard Fraunhofer pattern,~\cite{tinkham} we take into account the
dependence of the critical current $I_c(B_\|)$ on the parallel
component of the field. This is done by straightforwardly
generalizing the ballistic formula for the Josephson current,\cite{grajcar} 
in which we express the AR amplitude $a$ in
terms of the McMillan's Green's functions with the Doppler shift in
the InAs proximity layers. The dependence $I_c(B_\|)$ makes 
the Fraunhofer-like oscillations decay much faster
than the usual $1/B_\perp$ law, which is clearly seen in
Fig.~\ref{IC_B}. For $\alpha \leq 1.3^\circ$ no minimum is detected,
while it is still observed for a wider sample \#2a on the same
chip (not shown). The curve for $\alpha\approx 0$ resulted from a
careful maximization of $I_C$ at fixed field, since in this case,
evidently, no Fraunhofer minimum can be observed anymore. 
The fits (solid lines in Fig.~\ref{IC_B}) are obtained by varying 
the effective thickness of the Nb/InAs bilayer, $d_\textrm{eff}$. We
find $d_\textrm{eff}=d_N+d_S=50$~nm for sample \#2a (not shown) in
agreement with the geometry, while for sample \#1
$d_\textrm{eff}\approx 25$ nm. The reason for this discrepancy is unclear, 
since the excess current data in Fig.~\ref{Iexc} agree very
well. The deviation of the data for $\alpha\approx 0$ from
the theoretical $B_\perp=0$ curve is probably due to a weak perpendicular stray field 
of the Nb electrodes.

In conclusion, we have studied the IV-characteristics and critical current 
of ballistic Nb/InAs Josephson junctions in parallel and perpendicular magnetic
fields. The observed field dependence is much stronger than
anticipated from standard models and can be traced back
to the Doppler shift of Andreev levels due to diamagnetic
supercurrents in hybrid Nb/InAs contacts.
Several different aspects of the proximity effect are consistently
and nearly quantitatively described by our theoretical model. 

We thank J.~C.~Cuevas and N.~Shchelkachev for helpful discussions.
The work was supported by the DFG (STR 438-2 and GRK 638) and partially by MPI-PKS (G. T.).




\begin{thebibliography}{99}
\bibitem{conditions}
This is justified, since, for the parallel field, the London
penetration depth $\lambda_L\simeq 110$~nm is much larger than half
of the Nb/InAs bilayer thickness $(d_S+d_N)/2=25$~nm, while, for the
perpendicular field, the Pearl penetration depth
$\lambda_P=2\lambda_L^2/d_S\simeq 700$~nm is larger than half of the
Nb stripe width $W_S/2=500$~mm.
\end{thebibliography}
\end{document}